\documentclass[twocolumn,showpacs,aps,prl,superscriptaddress]{revtex4}

\usepackage{graphicx}
\usepackage{dcolumn}
\usepackage{epsfig}
\usepackage{amsmath}

\newcommand{\BaBarYear}    {08}
\newcommand{\BaBarNumber}  {23}
\newcommand{\SLACPubNumber} {13317}

\newcommand{\BaBarType}      {PUB}  

\RequirePackage{xspace}

\usepackage{relsize}
\def\babar{\mbox{\slshape B\kern-0.1em{\smaller A}\kern-0.1em
    B\kern-0.1em{\smaller A\kern-0.2em R}}}

\def\epem       {\ensuremath{e^+e^-}\xspace}

\def\q     {\ensuremath{q}\xspace}

\def\qqbar {\ensuremath{q\overline q}\xspace}
\def\u     {\ensuremath{u}\xspace}

\def\d     {\ensuremath{d}\xspace}

\def\s     {\ensuremath{s}\xspace}

\def\c     {\ensuremath{c}\xspace}

\def\piz   {\ensuremath{\pi^0}\xspace}

\def\pip   {\ensuremath{\pi^+}\xspace}
\def\pim   {\ensuremath{\pi^-}\xspace}

\def\Kbar  {\kern 0.2em\overline{\kern -0.2em K}{}\xspace}

\def\Kz    {\ensuremath{K^0}\xspace}
\def\Kzb   {\ensuremath{\Kbar^0}\xspace}
\def\KzKzb {\ensuremath{\Kz \kern -0.16em \Kzb}\xspace}
\def\Kp    {\ensuremath{K^+}\xspace}
\def\Km    {\ensuremath{K^-}\xspace}

\def\KpKm  {\ensuremath{\Kp \kern -0.16em \Km}\xspace}

\def\Dbar    {\kern 0.2em\overline{\kern -0.2em D}{}\xspace}

\def\Dz      {\ensuremath{D^0}\xspace}
\def\Dzb     {\ensuremath{\Dbar^0}\xspace}
\def\DzDzb   {\ensuremath{\Dz {\kern -0.16em \Dzb}}\xspace}
\def\Dp      {\ensuremath{D^+}\xspace}
\def\Dm      {\ensuremath{D^-}\xspace}

\def\DpDm    {\ensuremath{\Dp {\kern -0.16em \Dm}}\xspace}

\def\B       {\ensuremath{B}\xspace}
\def\Bbar    {\kern 0.18em\overline{\kern -0.18em B}{}\xspace}

\def\BB      {\ensuremath{B\Bbar}\xspace} 
\def\Bz      {\ensuremath{B^0}\xspace}
\def\Bzb     {\ensuremath{\Bbar^0}\xspace}
\def\BzBzb   {\ensuremath{\Bz {\kern -0.16em \Bzb}}\xspace}
\def\Bu      {\ensuremath{B^+}\xspace}
\def\Bub     {\ensuremath{B^-}\xspace}
\def\Bp      {\ensuremath{\Bu}\xspace}

\def\BpBm    {\ensuremath{\Bu {\kern -0.16em \Bub}}\xspace}

\def\BorBbar    {\kern 0.18em\optbar{\kern -0.18em B}{}\xspace}
\def\DorDbar    {\kern 0.18em\optbar{\kern -0.18em D}{}\xspace}
\def\KorKbar    {\kern 0.18em\optbar{\kern -0.18em K}{}\xspace}

\mathchardef\Upsilon="7107
\def\Y#1S{\ensuremath{\Upsilon{(#1S)}}\xspace}

\def\FourS {\Y4S}

\mathchardef\Deltares="7101
\mathchardef\Xi="7104
\mathchardef\Lambda="7103
\mathchardef\Sigma="7106
\mathchardef\Omega="710A

\def\Deltabar{\kern 0.25em\overline{\kern -0.25em \Deltares}{}\xspace}
\def\Lbar{\kern 0.2em\overline{\kern -0.2em\Lambda\kern 0.05em}\kern-0.05em{}\xspace}
\def\Sigbar{\kern 0.2em\overline{\kern -0.2em \Sigma}{}\xspace}
\def\Xibar{\kern 0.2em\overline{\kern -0.2em \Xi}{}\xspace}
\def\Obar{\kern 0.2em\overline{\kern -0.2em \Omega}{}\xspace}
\def\Nbar{\kern 0.2em\overline{\kern -0.2em N}{}\xspace}
\def\Xb{\kern 0.2em\overline{\kern -0.2em X}{}\xspace}

\def\mes        {\mbox{$m_{\rm ES}$}\xspace}

\def\DeltaE     {\mbox{$\Delta E$}\xspace}

\newcommand{\tev}{\ensuremath{\mathrm{\,Te\kern -0.1em V}}\xspace}
\newcommand{\gev}{\ensuremath{\mathrm{\,Ge\kern -0.1em V}}\xspace}
\newcommand{\mev}{\ensuremath{\mathrm{\,Me\kern -0.1em V}}\xspace}
\newcommand{\kev}{\ensuremath{\mathrm{\,ke\kern -0.1em V}}\xspace}
\newcommand{\ev}{\ensuremath{\mathrm{\,e\kern -0.1em V}}\xspace}
\newcommand{\gevc}{\ensuremath{{\mathrm{\,Ge\kern -0.1em V\!/}c}}\xspace}
\newcommand{\mevc}{\ensuremath{{\mathrm{\,Me\kern -0.1em V\!/}c}}\xspace}
\newcommand{\gevcc}{\ensuremath{{\mathrm{\,Ge\kern -0.1em V\!/}c^2}}\xspace}
\newcommand{\mevcc}{\ensuremath{{\mathrm{\,Me\kern -0.1em V\!/}c^2}}\xspace}

\def\invfb   {\ensuremath{\mbox{\,fb}^{-1}}\xspace}

\def\mus  {\ensuremath{\rm \,\mus}\xspace}

\def\mus        {\ensuremath{\,\mu{\rm s}}\xspace}    

\def\to                 {\ensuremath{\rightarrow}\xspace}

\def\pep2{PEP-II}

\def\gsim{{~\raise.15em\hbox{$>$}\kern-.85em
          \lower.35em\hbox{$\sim$}~}\xspace}
\def\lsim{{~\raise.15em\hbox{$<$}\kern-.85em
          \lower.35em\hbox{$\sim$}~}\xspace}

\def\deltat{\ensuremath{{\rm \Delta}t}\xspace}

\xspace

\newcommand{\epjBase}        {Eur.\ Phys.\ Jour.\xspace}
\newcommand{\jprlBase}       {Phys.\ Rev.\ Lett.\xspace}
\newcommand{\jprBase}        {Phys.\ Rev.\xspace}
\newcommand{\jplBase}        {Phys.\ Lett.\xspace}
\newcommand{\nimBaseA}       {Nucl.\ Instrum.\ Methods Phys.\ Res., Sect.\ A\xspace}

\newcommand{\npBase}         {Nucl.\ Phys.\xspace}

\newcommand{\epjc}      [1]  {\epjBase\ C~{\bf #1}}

\newcommand{\jpg}       [1]  {{J.\ Phys.\ {\bf G{\bf #1}}}}

\newcommand{\nima}      [1]  {\nimBaseA~{\bf #1}}

\newcommand{\npb}       [1]  {\npBase\ B~{\bf #1}}

\newcommand{\jpl}       [1]  {\jplBase\ {\bf #1}}
\newcommand{\jprl}      [1]  {\jprlBase\ {\bf #1}}
\newcommand{\jprd}      [1]  {\jprBase\ D~{\bf #1}}

\def\jetset74   {\mbox{\tt Jetset \hspace{-0.5em}7.\hspace{-0.2em}4}\xspace}

\newcommand{\e}      [1]   { {\ensuremath{ \times 10^{ {#1} } }}}
\def\br        {\ensuremath{ {\cal {B}}}}

\def\fish   {\ensuremath{\cal F}}

\def\mes   {\ensuremath{m_{\mathrm{ES}}}}

\def\dt   {\ensuremath{\Delta t}}

\def\coshel {\ensuremath{\cos(\theta) }}

\def\fish  {\ensuremath{\cal F}}

\def\dz   {\ensuremath{\Delta z}}

\def\rhoz   {\ensuremath{\rho^{0}}}

\def\qq   {\ensuremath{q \overline{q}}}
\def\bb   {\ensuremath{B \overline{B}}}
\def\ifb  {\ensuremath{\mathrm{fb}^{-1}}}

\def\ptrue   {\ensuremath{f_{L}}}

\def\piz {\ensuremath{\pi^0}}

\def\rhop {\ensuremath{\rho^+ }\xspace}

\def\fish    {\ensuremath{\cal F}}
\def\coshel  {\ensuremath{\cos\theta_{i}}}

\def\mvonetwo {\ensuremath{ m_{1,2}}}
\def\coshelonetwo {\ensuremath{\cos\theta_{1,2}}}

\def\phiphiyield{\ensuremath{-1.5^{+3.7}_{-2.9}}}

\def\kkkksamplesize{209}

\def\phirhoyield{\ensuremath{22.5^{+11.3}_{-9.7}}}

\def\kkppzsamplesize{3175}

\def\phirhozyield{\ensuremath{3.9^{+6.3}_{-4.4}}}
\def\fzfzyield{\ensuremath{-13.6^{+4.8}_{-3.5}}}
\def\phifzyield{\ensuremath{0.8^{+2.4}_{-1.4}}}

\def\kkppsamplesize{3949}

\def\ubztophirhoz {\ensuremath{3.3}}
\def\ubztophiphi {\ensuremath{2.0}}
\def\ubptophirhop {\ensuremath{30}}
\def\ubztophifzero {\ensuremath{3.8}}
\def\ubztofzerofzero {\ensuremath{2.3}}

\def\bfztophirhoz    {\ensuremath{0.9^{+1.3}_{-0.9}\pm 0.9}}
\def\bfztophiphi     {\ensuremath{-0.4^{+1.2}_{-0.9} \pm 0.3}}
\def\bfptophirhop    {\ensuremath{15^{+7}_{-6} \pm 9}}
\def\bfztophifzero   {\ensuremath{0.2^{+0.6}_{-0.3} \pm 0.3}}
\def\bfztofzerofzero {\ensuremath{-1.4^{+0.5}_{-0.4} \pm 1.5}}

\def\fzero {\ensuremath{f_{0}}\xspace}
\def\rhoz {\ensuremath{\rho^0}}
\def\rhop {\ensuremath{\rho^+}}

\def\bztophirhoz{\ensuremath{\Bz\rightarrow\phi\rhoz}}
\def\bztophiphi{\ensuremath{\Bz\rightarrow\phi\phi}}
\def\bptophirhop{\ensuremath{\Bp\rightarrow\phi\rhop}}
\def\bztophifzero{\ensuremath{\Bz\rightarrow\phi\fzero}}
\def\bztofzerofzero{\ensuremath{\Bz\rightarrow\fzero\fzero}}

\begin{document}

\preprint{\babar-PUB-\BaBarYear/\BaBarNumber} 
\preprint{SLAC-PUB-\SLACPubNumber} 

\begin{flushleft}
\babar-\BaBarType-\BaBarYear/\BaBarNumber \\
SLAC-PUB-\SLACPubNumber\\
\end{flushleft}


\title{
 \large \bf\boldmath Searches for $B$ meson decays to $\phi\phi$, $\phi\rho$, $\phi\fzero(980)$, and $\fzero(980)\fzero(980)$ final states
}

%
\author{B.~Aubert}
\author{M.~Bona}
\author{Y.~Karyotakis}
\author{J.~P.~Lees}
\author{V.~Poireau}
\author{E.~Prencipe}
\author{X.~Prudent}
\author{V.~Tisserand}
\affiliation{Laboratoire de Physique des Particules, IN2P3/CNRS et Universit\'e de Savoie, F-74941 Annecy-Le-Vieux, France }
\author{J.~Garra~Tico}
\author{E.~Grauges}
\affiliation{Universitat de Barcelona, Facultat de Fisica, Departament ECM, E-08028 Barcelona, Spain }
% ...
\collaboration{The \babar\ Collaboration}
\noaffiliation

\date{\today}

\begin{abstract}
We present the results of searches for \B\ decays to charmless final states 
involving $\phi$, $\fzero(980)$, and charged or neutral $\rho$ mesons.
The data sample corresponds to $384\e{6}$ \BB\ pairs collected with the
\babar\ detector operating at the PEP-II asymmetric-energy \epem\ collider
at SLAC.  We find no significant signals and determine the following 90\% 
confidence level upper limits on the branching fractions, including systematic
uncertainties:
$\br(\bztophiphi) < \ubztophiphi\e{-7}$,
$\br(\bptophirhop) < \ubptophirhop\e{-7}$,
$\br(\bztophirhoz) < \ubztophirhoz\e{-7}$,
$\br[\Bz\to\phi\fzero(980)] \times \br[\fzero(980)\to\pip\pim] < \ubztophifzero\e{-7}$, and
$\br[\Bz\to\fzero(980)\fzero(980)]\times \br[\fzero(980)\to\pip\pim]\times \br[\fzero(980)\to K^+ K^-] < \ubztofzerofzero\e{-7}$.
\end{abstract}


\maketitle

%
%

We report the results of searches for the decays 
$\Bz\to\phi\phi$, $\phi\rhoz$, $\phi\fzero(980)$, 
$\fzero(980)\fzero(980)$, and $\B^\pm\to\phi\rho^\pm$~\cite{fz980}
using data collected with the \babar\ detector.
The $\bztophiphi$ decay is an OZI suppressed process with an expected
branching fraction in the range (0.1 to 3) $\e{-8}$ in the Standard 
Model (SM)~\cite{shalom,lu2,beneke}.
The decays $\Bz\to\phi\rhoz$ and $\Bp\to\phi\rho^+$ are pure 
$b\to d$ loop processes; the expected branching fractions for these 
modes range from ($2$ to $7$)$\times 10^{-8}$~\cite{zou,lu,li,bao,gronau}.
The presence of new physics (NP) would give rise to additional amplitudes that could
enhance the branching fractions for
these decay modes relative to the SM predictions~\cite{shalom,lu,lu2}.
The branching fraction for $\bztophiphi$ could be enhanced to $10^{-7}$~\cite{shalom},
and the branching fractions for $\B\to\phi\rho$ decays could be enhanced by 
20\%~\cite{bao} in the presence of NP.  We are not aware of branching fraction
predictions for $\Bz\to\phi\fzero$ and $\Bz\to\fzero\fzero$.

The \B\ decays to $\phi\phi$ and $\phi\rho$ are 
complicated by the presence of one amplitude with longitudinal polarization and
two amplitudes with transverse polarization.
The fraction of longitudinally polarized events is denoted by $\ptrue$. Integrating over 
the angle between the vector meson decay planes, the angular distribution 
$(1/\Gamma) d^2\Gamma / d\cos\theta_1 d\cos\theta_2$ is
\begin{eqnarray}
 \frac{9}{4} \left[f_L \cos^2\theta_1 \cos^2\theta_2 + \frac{1}{4}(1-\ptrue) \sin^2\theta_1 \sin^2\theta_2 \right],\label{eq:polarisation}
\end{eqnarray}
where the indices $1,2$ label the two vector mesons in the final state, and the helicity angles 
$\theta_{1,2}$ are the angles between the direction opposite to that of the $B^0$ ($B^+$)
and the $K^+$ or $\pi^+$ (\piz) momentum in the $\phi$ or $\rhoz$ ($\rho^+$) rest
frame.  We define the angles $\theta_{1,2}$ for \fzero\ mesons in an analogous way. 
The expected values of $\ptrue$ range from $0.6$ to 0.8~\cite{li,lu,lu2,beneke}
for $\Bz\to\phi\phi$, $\phi\rhoz$, and $\B^\pm\to\phi\rho^\pm$.  The presence
of NP could lead to enhancements of the transverse polarization amplitudes~\cite{shalom,lu,lu2}.

The current upper limit on the $\bztophiphi$ branching fraction,
obtained from a data sample of 82 \invfb, is $1.5\e{-6}$~\cite{babarphiphi}.
The upper limits on  $\Bz\to\phi\rhoz$ and $\Bp\to\phi\rho^+$,
determined using 3.1 \invfb of data, are  $1.3\e{-5}$ and $1.6\e{-5}$~\cite{cleovv},
respectively.  Using a data sample of 349 \invfb, \babar\ recently reported
an upper limit of $1.6\e{-7}$ for $\Bz\to\fzero\fzero$~\cite{babarrho0rho0}.  This last result relies on
the assumption that the $\fzero\to \pi^+\pi^-$ branching fraction is 100\%.  In this
analysis, we make the complimentary assumption that one $\fzero$ decays to
$\pi^+\pi^-$ and the other to $K^+K^-$ and search for $\Bz\to\fzero\fzero$ in a 
cleaner final state than Ref.~\cite{babarrho0rho0}.  All these limits correspond to a confidence
level (C.L.) of 90\%.

The results presented here are based on an integrated luminosity of 349\ifb, 
corresponding to $(384\pm 4)$ million \BB\ pairs.  These data were 
recorded at the \FourS resonance with a center-of-mass (CM) energy
$\sqrt{s}$ = 10.58 $\gev$. 
The \babar\ detector is described in detail elsewhere~\cite{babarnim},
and is situated at the interaction region of the \pep2\ asymmetric energy 
$\epem$ collider located at the Stanford Linear Accelerator Center (SLAC).
We use Monte Carlo (MC) simulated events generated using the GEANT4
based~\cite{ref:geant} \babar\ simulation.

%
%

Photons are reconstructed from localized deposits of energy greater than 50\mev\ in the 
electromagnetic calorimeter
that are not associated with a charged track.  We require 
$\gamma$ candidates to have a lateral
shower profile~\cite{ref:lat} that is consistent with the expectation for photons.
\piz\ candidates are reconstructed from two $\gamma$ candidates with invariant
mass $0.10 < m_{\gamma\gamma} < 0.16$ \gevcc.

We use information from the vertex detector, drift chamber and detector of internally 
reflected Cherenkov light to select charged tracks that are consistent with
kaon or pion signatures in the detector~\cite{s2bprd}.
We reconstruct $\phi$ ($\rhoz$) candidates from pairs of oppositely charged kaon (pion)
candidates with invariant mass $0.99<m_{KK}<1.05 \gevcc$ ($0.55 < m_{\pi\pi} < 1.05 \gevcc$).
For $\rhoz$ candidates we require the helicity angles to satisfy 
$|\coshel|<0.98$ since signal efficiency falls off near $|\coshel| = 1$.
Charged $\rho$ candidates are reconstructed from a charged track consistent
with the pion signature and a \piz\ candidate. The invariant mass 
$m_{\pi\piz}$ of the $\rho^+$ candidate is required to lie between
 0.5 and 1.0 \gevcc. We also require that the helicity angles satisfy $-0.8<\coshel<0.98$
as signal efficiency is asymmetric because of the \piz\ meson, and falls off near 
$\coshel = \pm 1$, and background peaks near $-1$.
We select $\fzero$ candidates from two charged tracks that are both either consistent 
with the kaon or the pion signature in the detector.  
We apply the same selection criteria to $\fzero\to\pip\pim$
candidates as for $\rhoz$ mesons. Similarly we apply the same
selection criteria to $\fzero\to K^+K^-$ candidates as for $\phi$ mesons
as the minimum $m_{KK}$ we can reconstruct in the detector
is 0.99 \gevcc.

We reconstruct signal \B\ candidates ($B_{\mathrm{rec}}$) from combinations of two $\phi$ mesons, one $\phi$ and 
one $\rho$ or $\fzero$, and two $\fzero$ mesons.  The $\fzero\fzero$ mode is required to have one $\fzero$ 
decaying into $\pi^+\pi^-$, and the other decaying into $K^+K^-$.  We require the $\fzero$ in $\phi\fzero$ 
to decay into $\pi^+\pi^-$.

We use two kinematic variables, $\mes$ and \DeltaE, in order to isolate the signal:
$\mes=\sqrt{(s/2 + {\mathbf {p}}_i\cdot {\mathbf {p}}_B)^2/E_i^2- {\mathbf {p}}_B^2}$ is the beam-energy substituted mass and
$\DeltaE = E_B^* - \sqrt{s}/2$ is the difference between the \B\ candidate energy
and the beam energy in the $e^+e^-$ CM frame. Here the $B_{\mathrm{rec}}$ momentum
${\mathbf {p}_B}$ and four-momentum of the initial state $(E_i, {\mathbf
{p}_i})$ are defined in the laboratory frame, and $E_B^*$ is the $B_{\mathrm{rec}}$ energy in
the $e^+e^-$ CM frame.
The distribution of \mes\ (\DeltaE) peaks at the \B mass (near zero) for signal events
and does not peak for background.
We require $\mes > 5.25 \gevcc$.  For the $\phi\phi$ final state we require $|\DeltaE| < 0.15 \gev$.
To reduce background from non-signal $\B$ meson decays we apply the more stringent 
cut of $-0.07 < \DeltaE < 0.15$ \gev for all other modes.

The angle in CM frame between the thrust axis
of the rest of the event (ROE) and that of the \B\ candidate is required to satisfy $|\cos(\theta_{T\!B, T\!R})|<0.8$
in order to reduce background from  \epem\to\qqbar ($\q = \u,\d,\s,\c$) continuum events.
The variable $|\cos(\theta_{T\!B, T\!R})|$ is strongly peaked near 1 for \qq\ events, whereas
\bb\ events are more isotropic because the \B\ mesons are produced close to the kinematic threshold.
Additional separation between signal and continuum events is obtained by combining several kinematic and 
topological variables into a Fisher discriminant $\fish$, which we use in the maximum-likelihood fit described below. 
The variables $|\cos(\theta_{T\!B, T\!R})|$, $|\deltat|/\sigma(\deltat)$, $|\cos(\theta_{B, Z})|$, 
$|\cos(\theta_{T\!B, Z})|$,
and the output of a multivariate tagging algorithm~\cite{tag} are used as inputs to $\fish$. 
The time interval \dt\ is calculated from the measured separation distance \dz\ between
the decay vertices of $B_{\mathrm{rec}}$ and the other \B\ in the event ($B_{\mathrm{ROE}}$) 
along the beam axis ($z$). The vertex of $B_{\mathrm{rec}}$ is reconstructed from the 
tracks that come from the signal candidate; the vertex of $B_{\mathrm{ROE}}$ is reconstructed 
from tracks in the ROE, with constraints from 
the beam spot location and the $B_{\mathrm{rec}}$ momentum. The uncertainty on the measured value
of \dt\ is $\sigma(\deltat)$.
The variable $\theta_{B, Z}$ is the angle between the direction 
of $\B_{\mathrm{rec}}$ and the $z$ axis in the CM frame. This variable follows a sine squared distribution 
for \bb\ events, whereas it is almost uniform for \qq.
The variable $\theta_{T\!B, Z}$ is the angle between the \B\ thrust direction and the $z$ axis in the 
laboratory frame. 

The decay modes studied are classified into three groups according to the final state particles:
(i) \bztophiphi, (ii) \bptophirhop, and (iii) \bztophirhoz, \bztophifzero, and \bztofzerofzero.
We find that 6\% of events for the mode in group (ii) and 3\% of events for the modes in group 
(iii) have more than one candidate that passes our selection criteria.  For such events we 
retain the candidate with the smallest $\chi^2$ for the $B_{\mathrm{rec}}$ vertex for use in 
the fits described below.
The numbers of selected candidates are given in Table~\ref{tab:results}.

%
%
The dominant background for all modes comes from continuum events. The yield of this
background component is determined from the fit to data.
The dominant \B\ backgrounds for group (i) are $\Bz\to\phi K^{*0}$ and $\fzero K^{*0}$,
which are estimated to contribute 1.4 and 0.6 events to the data, respectively. The \B\ backgrounds for 
group (ii) are events from \B\ decays to final states including charm and
$\B^+ \to \phi K^{*+}$.  These are estimated to contribute 107 and 5.5 events to the data. 
The \B\ backgrounds for group (iii) are
events from \B\ decays to final states including charm, $\Bz$ decays to $\phi K^{*0}$,
$\fzero K^{*0}$, $\phi K^{*0}_2(1430)$, and $\B^+$ decays to
$\phi K^{+}$ and $\phi K^{*+}$ estimated to contribute 249, 25.9, 9.1, 2.3, 
4.7, and 1.8 events to the data.  The branching fractions for the 
\B\ backgrounds are taken from Ref.~\cite{PDG2006}, except for $\Bz\to\fzero K^{*0}$,
which has not yet been measured, and $\phi \rho^{+}$ where we use the results obtained here.  
The current upper limit on the $\Bz\to\fzero K^{*0}$ branching fraction is $4.3\e{-6}$ and we 
assume a branching fraction of $(2 \pm 2)\e{-6}$.

%
%
We obtain yields for each mode from extended unbinned maximum likelihood (ML) fits with 
the input observables $\mes$, $\DeltaE$, and $\coshelonetwo$. In addition, for 
all modes except $\phi\phi$, we include $\mvonetwo$ and  $\fish$ in the 
likelihood, where $\mvonetwo$ is $m_{\pi\pi}$ or $m_{KK}$ for the 
$\phi$, $\rho$ or $\fzero$ candidates. A total of three fits are performed, 
one for each group of signal modes.
We include event hypotheses for signal events and 
the aforementioned backgrounds in each of the fits. 
For each event $i$ and hypothesis $j$, the likelihood function is
\begin{equation}
{\cal L}= \frac{e^{-\left(\sum n_j\right)}}{N!} \prod_{i=1}^N \left[\sum_{j=1}^{N_j} n_j  {\cal P}_j ({\bf x}_i)\right]\ , \nonumber
\end{equation}
where  $N$ is the number of input events, $N_j$ is the number of hypotheses, $n_j$ is the number of events for hypothesis $j$ and     
${\cal P}_j ({\bf x}_i)$ is the corresponding probability density function (PDF) evaluated
for the observables ${\bf x}_i$ of the $i^{th}$ event.
The correlations between input observables are small and are assumed to be negligible. 
Possible biases due to residual correlations are evaluated as described below.
We compute the combined PDFs ${\cal P}_j ({\bf x}_i)$ as the product of PDFs for each of the 
input observables.
These combined PDFs are used in the fit to the data.

\begin{table*}[!ht]
\caption{Number of events $N$ in the data sample, signal yield $\cal{Y_{S}}$ (corrected for fit bias), 
fit bias, detection efficiency $\epsilon$, 
daughter branching fraction product ($\prod{\cal{B}}_i$), 
significance $\sigma$ (including additive systematic uncertainties, taken to be zero if the fitted yield is negative), 
measured branching fraction where the first error is statistical, and the second 
systematic (see text), and the 90\% C.L. upper limit on this branching fraction (including systematic uncertainties).
For $\B$ decays to $\phi\phi$ and $\phi\rho$, two efficiencies are reported, 
one for longitudinally and one for transversely polarized events.
The reported branching fractions for $\phi \fzero$ and $\fzero\fzero$ are product branching fractions that are
not corrected for the probability of $\fzero$ decaying into $\pip\pim$ or $K^+K^-$.}
\label{tab:results}
\begin{tabular}{cccccccccc}\hline\hline
\vspace{-0.25cm}
\\
\phantom\  Group \phantom\ & \phantom\  $N$ \phantom\  & \phantom\  Mode \phantom\  &  \phantom\  $\cal{Y_{S}}$  \phantom\ &  \phantom\  Bias\phantom\   & \phantom\  $\epsilon$(\%)  \phantom\ &  \phantom\ $\prod{\cal{B}}_i$(\%)  \phantom\ &  \phantom\ $\sigma$  \phantom\ &  \phantom\ \br(\e{-7})  \phantom\ &  \phantom\ UL($\times 10^{-7}$)  \phantom\ \\[2pt] \hline 
\vspace{-0.25cm}
\\
(i)   & \kkkksamplesize  &{\boldmath{$\phi\phi$}}    & \phiphiyield  & $-0.4\pm 0.2$ & 40.4 [28.7]   & $24.3\pm 1.2$     & $0.0$   & $\bfztophiphi$ & $<${\bf{\ubztophiphi}}    \\[2pt]\hline\noalign{\vskip1pt}
(ii)  & \kkppzsamplesize &{\boldmath{$\phi\rhop$}}   & \phirhoyield  & $+2.3\pm 1.1$ & 5.7 [9.8]     & $49.3\pm 0.6$     & $2.2$ & $\bfptophirhop$ & $<${\bf{\ubptophirhop}}    \\[2pt]\hline\noalign{\vskip1pt}
(iii) &\kkppsamplesize   & {\boldmath{$\phi\rhoz$}}  & \phirhozyield & $+0.8\pm 0.4$ & 24.1 [26.5]   & $49.3\pm 0.6$     & $1.0$ & $\bfztophirhoz$ & $<${\bf{\ubztophirhoz}}    \\[2pt]
      &                  &{\boldmath{$\phi\fzero$}}  & \phifzyield   & $-1.7\pm 0.5$ & 22.1          & $\ldots$          & $0.0$ & $\bfztophifzero$ & $<${\bf{\ubztophifzero}}    \\[2pt]
      &                  &{\boldmath{$\fzero\fzero$}}  & \fzfzyield    & $-1.8\pm 0.5$ & 25.5          & $\ldots$           & $0.0$ & $\bfztofzerofzero$ & $<${\bf{\ubztofzerofzero}}    \\[3pt] \hline\hline
\end{tabular}
\end{table*}

%
%
For $B$ decays to $\phi\phi$ and $\phi\rho$, the \mes\ distribution is parametrized with 
the sum of a Gaussian and a Gaussian with a low-side exponential component.  The
\DeltaE distribution is described by the sum of two Gaussian distributions, and the 
$\coshelonetwo$ distributions are described by Eq.~(\ref{eq:polarisation}) multiplied by an acceptance function.  The acceptance function 
is a polynomial for all $\coshelonetwo$, with the exception of the $\rho^+$ helicity angle 
distribution for longitudinally polarized $\phi\rho^+$, which uses a polynomial multiplied by 
the sigmoid function $1/(1+\exp[\alpha(\coshelonetwo+\beta)])$, where the parameters $\alpha$ and $\beta$ are
determined from MC simulated data. For the $\phi\rho$ final states we use a Gaussian to 
describe the \fish\ distribution, and the sum of a relativistic Breit-Wigner (BW) with two Gaussians for $\mvonetwo$. 
The continuum background \mes\ distribution is described by an ARGUS function~\cite{ref:argus}.
We parameterize the continuum \DeltaE\ distribution using a second-order polynomial and use 
polynomials to describe $\coshelonetwo$. Where appropriate, we parameterize the \fish\ distributions for 
the continuum background using a Gaussian, and we parameterize the $\mvonetwo$ distributions 
using the sum of a BW and a polynomial.
We use smoothed histograms of MC simulated data as the PDFs for all other signal and background modes.
We generate $\Bz\to\phi\fzero$ assuming that the $\phi$ is longitudinally polarised, and we use
phase space distributions for $\Bz\to\fzero\fzero$.
Before fitting the data, we validate the fitting procedure using 
the methods described in Ref.~\cite{rhorhoprd}.  We determine a bias correction
on our ability to correctly determine the signal yield using ensembles of 
simulated experiments generated from samples of MC simulated data for the signal
and exclusive backgrounds and from the PDFs for the other backgrounds.

%
%
Our results are summarized in Table~\ref{tab:results} where we show the measured yield, 
fit bias, efficiency, and the product of daughter branching fractions for each decay mode. 
We compute the branching fractions from the fitted signal event 
yields corrected for the fit bias, reconstruction efficiency, daughter branching fractions, and the number of 
produced \B mesons, assuming equal production rates of charged and neutral \B pairs. 
As we do not know the value of \ptrue\ for the $\phi\phi$ and $\phi\rho$ modes, we fit
the data for different physically allowed values of \ptrue\ in steps of 0.1.
We find no evidence for any of the signal modes and calculate 90\% C.L. branching 
fraction upper limits $x_{UL}$ such that
$\int_0^{x_{UL}}L({\cal{Y_{S}}},\ptrue) d{\cal{Y_{S}}} / \int_{0}^{+\infty}L({\cal{Y_{S}}},\ptrue) d{\cal{Y_{S}}} =0.9$,
where $L({\cal{Y_{S}}}, \ptrue)$ is the likelihood as a function of signal yield ${\cal{Y_{S}}}$ and \ptrue\ 
multiplied by a uniform prior. We report the most conservative (largest) upper limits for each mode, for which 
$\ptrue=0.5$, 0.7, and 0.2 for groups (i), (ii), and (iii), respectively.
The central values of the branching fractions given in Table~\ref{tab:results}
correspond to these values of \ptrue. Figure~\ref{fig:data} shows the \mes\ 
distributions in subsamples of the data where $|\Delta E| < 0.05$ \gev\ for 
$\bptophirhop$, and $|\Delta E| < 0.025$ \gev\ for all other modes.
\begin{figure}[!ht]
\begin{center}
\resizebox{7.cm}{!}{
\includegraphics{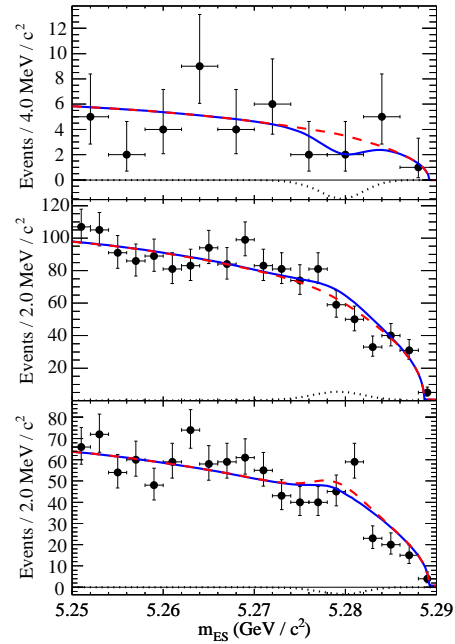}
}
\caption{(color online) Signal-enhanced distributions of \mes\ in 
data, with a projection of the fitted likelihood for (top) \bztophiphi, 
(middle) \bptophirhop, and (bottom) \bztophirhoz, \bztophifzero, and \bztofzerofzero.
The solid line represents the total PDF, the dotted line represents signal,
and the dashed line represents the sum of continuum and $\B$ backgrounds.}\label{fig:data}
\end{center}
\end{figure}

%
%
We estimate the systematic uncertainty related to the parameterization of the PDF 
by varying each parameter 
by its estimated uncertainty, and by substituting smoothed histograms by un-smoothed ones.  
The total contribution of all variations in signal yields, when added in 
quadrature, gives an error between $0.2$ and $5.6$ events, depending on the mode.
We account for possible differences between data and MC events from studies of 
a control sample of $B\to D\pi$ events, yielding an uncertainty of $0.1$ to
$12.2$ events depending on the mode. The 
uncertainty from fit bias is taken to be half the correction listed in 
Table~\ref{tab:results}. Incorporating the statistical uncertainty of 
the bias has a negligible effect.
The uncertainty on $B$-daughter branching fractions is 
in the range ($1.2$ to $4.9$)\%~\cite{PDG2006}.
The modes in group (iii), $\phi\rho^0$, $\phi\fzero$, and $\fzero\fzero$ have systematic
uncertainties from the \fzero\ lineshape~\cite{fzsyst} of 0.2, 3.1, and 15.9 events,
respectively.
The mode $\Bp\to \phi\rho^+$ has a fractional systematic uncertainty of 
3.0\% from the reconstruction efficiency of $\piz$ mesons.
Other sources of systematic errors are track reconstruction 
efficiency [($2.4 - 3.2$)\%], uncertainty on the number of $\B$ meson pairs (1.1\%), 
particle identification efficiency (3.5\%), and differences between data and MC efficiencies
related to the cut on the vertex $\chi^2$ (0.6\%).

Assuming isospin is conserved in $\fzero \to hh$ decays, where $h=\pi, K$, we correct for factors of 
${\cal{B}}(\fzero\to hh) / {\cal{B}}(\fzero\to h^+h^-)$ , to obtain
the product branching fraction upper limits of ${\cal{B}}(\Bz\to\phi\fzero)\times {\cal{B}}(\fzero\to\pi\pi) < 5.7\e{-7}$,
and ${\cal{B}}(\Bz\to\fzero\fzero)\times {\cal{B}}(\fzero\to\pi\pi)\times {\cal{B}}(\fzero\to KK)<6.9\e{-7}$ at 90\% C.L.

%
%
In summary we have performed searches for the decays $\Bz\to\phi\phi$, $\phi\rhoz$, $\phi\fzero$, 
$\fzero\fzero$, and $\B^\pm\to \phi\rho^\pm$ and place upper limits on these modes.
The upper limit on $\Bz\to \phi\phi$ reported here can be used to constrain 
possible NP enhancements suggested in Ref.~\cite{shalom}. 

We are grateful for the excellent luminosity and machine conditions
provided by our \pep2\ colleagues, 
and for the substantial dedicated effort from
the computing organizations that support \babar.
The collaborating institutions wish to thank 
SLAC for its support and kind hospitality. 
This work is supported by
DOE
and NSF (USA),
NSERC (Canada),
CEA and
CNRS-IN2P3
(France),
BMBF and DFG
(Germany),
INFN (Italy),
FOM (The Netherlands),
NFR (Norway),
MES (Russia),
MEC (Spain), and
STFC (United Kingdom). 
Individuals have received support from the
Marie Curie EIF (European Union) and
the A.~P.~Sloan Foundation.



\begin{thebibliography}{99}
\bibitem{fz980}
Throughout this paper, charge conjugation states are implied, and when we refer to \fzero, we mean specifically $f_0(980)$.
\end{thebibliography}
\end{document}